\documentclass[epj]{svjour}
\usepackage{graphics}
\usepackage{graphicx}
\usepackage{multirow}
\usepackage{amsmath,amssymb,amsfonts}
\usepackage{mathrsfs}
\usepackage{xcolor}
\usepackage{textcomp}
\usepackage{manyfoot}
\usepackage{booktabs}
\usepackage{algorithm}
\usepackage{algorithmicx}
\usepackage{algpseudocode}
\usepackage{listings}
\usepackage{tabularx}
\usepackage{hyperref}
\newcommand*{\tr}{\ensuremath{\mathrm{Tr}}}

\newcommand*{\cell}[1]{\begin{tabular}[c]{@{}l@{}}#1\end{tabular}}

\begin{document}
\title{HFB3: an axial HFB solver with Gogny forces using a 2-center HO basis (C++/Python)}

\author{N.~Dubray\inst{1,2}     \and
        J.-P.~Ebran\inst{1,2}   \and
        P.~Carpentier\inst{1,2} \and
        M.~Frosini\inst{3}      \and
        A.~Zdeb\inst{1,2,4}     \and
        N.~Pillet\inst{1,2}     \and
        J.~Newsome\inst{1,2}    \and
        M.~Verri\`ere\inst{1,2} \and
        G.~Accorto\inst{1,2}    \and
        D.~Regnier\inst{1,2}
}

\institute{
        CEA, DAM, F-91297, Arpajon, France                                                                                    \and
        Laboratoire Mati\`ere en Conditions Extr\^emes, Universit\'e Paris-Saclay, CEA, 91680, Bruy\`eres-le-Ch\^atel, France \and
        CEA, DES, F-13108, Saint-Paul-Lez-Durance, France                                                                     \and
        Department of Theoretical Physics, Institute of Physics, Maria Curie--Sk\l odowska University, Lublin, Poland
}

\date{Received: date / Revised version: date}
%
\abstract{
The HFB3 program solves the axial nuclear Hartree-Fock-Bogoliubov (HFB) equations using bases formed by either one or two sets of deformed Harmonic Oscillator (HO) solutions with D1-type and D2-type Gogny effective nucleon-nucleon interactions.
Using two sets of HO solutions shifted along the $z$-axis (2-center basis) allows to accurately describe highly elongated nuclear systems while keeping a moderate basis size, making this type of basis very convenient for the description of the nuclear fission process.
For the description of odd-even and odd-odd systems, the equal-filling-approximation is used.
Several observables can be calculated by the program, including the mean values of the multipole moments, nuclear radii, inertia tensors following Adiabatic Time-Dependent Hartree-Fock-Bogoliubov (ATDHFB) or Generator Coordinate Method (GCM) prescriptions, local and non-local one-body densities, local and non-local pairing densities, some fission fragment properties, etc.
The program can ensure that the mean values associated with some specific operators take pre-defined values (constraints). Such constraints can be set on the usual multipole moments (for protons, neutrons or total mass).
This program can be used as a monoprocess and monothreaded CLI executable, or through full-featured Python bindings (available through the Python Package Index PyPI).
\PACS{
  {21.60.Jz}{Self-consistent field calculations in nuclear structure} \and
      {24.75.+i, 25.85.-w}{Nuclear fission}   \and
      {21.10.Gv}{Nuclear deformation, nucleon distribution}   \and
      {13.75.Cs, 13.85.-t}{Nucleon-nucleon interactions}
     } 
} 


\maketitle

\noindent
{\bf Program Summary and Specifications}\\
\begin{small}
\noindent
{Program title:} HFB3\\
{Licensing provisions:} GPLv3\\
{Programming language:} C++ and Python\\
  {Repository:} \url{https://www.github.com/cea-phynu/hfb3}\\
{Description of problem:} The structure of an atomic nucleus can be described using the Energy Density Functional (EDF) approach.
The HFB3 program implements one such description, namely the Hartree-Fock-Bogoliubov theory.
The energy density functional used comes from a finite-range effective two-body nucleon-nucleon interaction of Gogny type.\\
{Method of solution:} The HFB3 program solves the nuclear Hartree-Fock-Bogoliubov equations using a deformed axial HO basis or the union of two sets of $z$-shifted deformed axial HO bases.
Constraints can be placed on specific multipole moments, allowing the calculation of potential energy surfaces of any dimension.
Specific numerical techniques are used to ensure a fast HFB convergence and a good minimization success rate.\\
{Additional comments:} The HFB3 program can be used as a simple CLI executable, or through the included Python bindings.
  The mandatory dependencies are specified in the \texttt{README.md} file and are publicly available under open-source licenses. \\
\end{small}

\section{Introduction}\label{sec1}

Among microscopic approaches to the nuclear many-body problem, the Energy Density Functional (EDF) method is arguably the one maintaining an optimal
compromise between global accuracy and feasibility of computational
cost, particularly in the heavy-mass region where the framework of EDFs
still remains unrivaled~\cite{Schunck19}. To describe an A-nucleon system, the EDF method postulates the existence of an effective Hamiltonian $H_\textrm{EDF}$ acting in a subspace of the A-body Hilbert space $\cal{H}_A$ and yielding the same low-energy observables of the original nuclear Hamiltonian which operates within $\cal{H}_A$. Taking this subspace as the one spanned by Bogoliubov vacua leads to the Hartree-Fock-Bogoliubov (HFB) realization of the EDF.

Within the EDF method, the treatment of nucleon correlations at the HFB level makes it possible to already provide a first level of description of a bunch of structure and fission properties\cite{Ben03,Sch16}. Namely the bulk of so-called dynamical correlations is implicitly taken into account by the empirical effective Hamiltonian $H_\textrm{EDF}$ while so-called static correlations, responsible for collective behaviors such as deformation, clustering or superfluidity, are explicitly included by allowing the HFB state to spontaneously break the symmetries of the nuclear Hamiltonian (cf. chapter 1 in \cite{Schunck19}).

The HFB method allows for a good description of the first low-energy excited states using quasi-particle excitations and/or beyond-mean-field methods such as GCM, RPA, QRPM, etc... Those structure studies correspond to moderate nuclear deformations and can be done in an efficient way with the HFB3 solver using 1-center or 2-center bases.

Constrained HFB calculations can also be used as a first step for some microscopic descriptions of the nuclear fission~\cite{Reg19}, involving highly deformed configurations. For such calculations, using 2-center bases allows to keep reasonable numerical cost and calculation time compared to equivalent 1-center HFB solvers.

\section{Formalism}
 \subsection{HFB formalism}


The effective nuclear Hamiltonian in second quantization is assumed to be of the form:
\begin{equation}\label{4.1}
\widehat{H} = \sum_{pq} \;
\langle p  \vert \widehat{t} \vert q \rangle \;  b_{p}^\dagger b_{q}  +
\frac{1}{4} \sum_{pqrs}   \; \langle p r  \vert \widehat{v}_{12}^{(a)}
\vert  q s \rangle  \;  b_{p}^\dagger b_{r}^\dagger b_{s} b_{q}.
\end{equation}
The kinematic energy operator $\widehat{t}$ is assumed to include the one-body center-of-mass correction:
\begin{equation}\label{4.2}
\widehat{t} = \left(1 - \frac{1}{A}\right)\frac{\widehat{p}^{2} }{2 M},
\end{equation}
and $\widehat{v}_{12}^{(a)}$ is an antisymmetrized two-body nuclear effective interaction:
\begin{equation}\label{4.3}
 \widehat{v}_{12}^{(a)} = \widehat{v}_{12}  \;
( 1 - \widehat{P}_{r} \widehat{P}_{\sigma} \widehat{P}_{\tau} ),
\end{equation}
where $\widehat{P}_{r}$, $\widehat{P}_{\sigma}$ et $\widehat{P}_{\tau}$
are respectively the exchange operators for space, spin and isospin variables.
It is assumed that the effective interaction depends on the nuclear density
$\rho(\vec{r})$ and that it includes the Coulomb interaction as well as the two-body center-of-mass correction:
\begin{align}\label{4.4}
  \langle p r  \vert \widehat{v}_{12}^{(a)} \vert  q s \rangle &=
\langle p r  \vert \widehat{v}_{NN}^{(a)} \vert  q s \rangle
  + \langle p r  \vert \widehat{v}_{C}^{(a)} \vert  q s \rangle \nonumber\\
  &+
\frac{1}{A M} \left( \langle p \vert \widehat{p} \vert s \rangle
\langle r \vert \widehat{p} \vert q \rangle  -
\langle p \vert \widehat{p} \vert q \rangle
\langle r \vert \widehat{p} \vert s \rangle \right).
\end{align}
Applying Wick's theorem to (\ref{4.1}) and using the time-reversal
symmetry properties, the total energy
${\cal E} \equiv  \langle \widetilde{0}  \vert \widehat{H}  \vert \widetilde{0}
 \rangle$ is written as:
\begin{eqnarray}
  {\cal E} & = & \sum_{pq>0} \;
    \langle p  \vert \widehat{t} \vert q \rangle \; \left( \rho_{qp}   +
       \overline{\rho}_{qp}  \right)
       \nonumber \\
  &  +&  \sum_{pqrs>0}   \;  \left[ 
  \langle p r  \vert \widehat{v}_{12}^{(a)}  \vert  q s \rangle  \;
         \frac{1}{2} \left(  \rho_{qp} \rho_{sr} +
         \overline{\rho}_{qp} \overline{\rho}_{sr}  \right) \right. \nonumber \\
  &&+ \left. \langle p \overline{r}  \vert \widehat{v}_{12}^{(a)}  \vert
          q \overline{s} \rangle  \;  \rho_{qp} \overline{\rho}_{rs} \right]
           \nonumber  \\
& + &  \sum_{pqrs>0}
   \langle  \overline{p} r \vert \widehat{v}_{12}^{(a)} \vert
            \overline{q} s  \rangle  \;   \kappa_{rp}  \kappa_{sq},
     \label{4.7}
\end{eqnarray}
where $\vert \widetilde{0} \rangle$ is the Bogoliubov vacuum which is assumed time-reversal invariant, and
\begin{align}
\left\{\begin{array}{l}
\rho_{pq}=\langle  \widetilde{0} \vert b^\dagger_p b_q\vert \widetilde{0} \rangle,\quad
\overline{\rho}_{pq}=\langle  \widetilde{0} \vert b^\dagger_{\overline{p}}
 b_{\overline{q}}\vert \widetilde{0} \rangle, \\
\kappa_{pq}=\langle  \widetilde{0} \vert b_p b_{\overline{q}}\vert \widetilde{0} \rangle=\langle  \widetilde{0} \vert b^\dagger_{\overline{q}}b^\dagger_p \vert \widetilde{0} \rangle,\\
\end{array}\right.
  \label{3.26}
\end{align}
are a simplified writting of the one-body density and pairing tensor for $p,q>0$. The $\overline{p}$ and $\overline{q}$ indices denote time-reversal partners of the $p$ and $q$ indices.\\

\noindent Then, one defines the functional:
\begin{equation}\label{4.9}
{\cal F}[\rho,\overline{\rho},\kappa]  =  {\cal E}[\rho,\overline{\rho},\kappa] - \sum_{\tau} \; \mu_{\tau}
\langle \widetilde{0} \vert \widehat{C}_{\tau}  \vert \widetilde{0}  \rangle,
\end{equation}
where $\hat{C}_{\mu}$ are constraints associated with multipole deformations, particle numbers or 1-body geometry properties (neck operator, fragment mass asymmetry).

\noindent At this point, it's convenient to introduce the generalized density matrix $\mathcal R$ which is symmetric when time-reversal is preserved:
\begin{equation}\label{3.10}
\displaystyle
 \mathcal R \ = \
  \left( \begin{array}{cc} \rho & -\kappa \\
  -\kappa^T & I - \rho  \end{array}  \right),
\end{equation}
which is idempotent:
\begin{equation}\label{3.8}
\displaystyle
 \mathcal R^2 \ = \mathcal R.
\end{equation}
In order to take into account this last relation, it is usual to further generalize the
functional to be minimized by applying the variational principle to the expression:
\begin{equation}\label{4.10}
{\cal G}[\rho,\overline{\rho},\kappa] = {\cal F}[\rho,\overline{\rho},\kappa] -
\tr \Lambda ( R^{2} - R).
\end{equation}
Applying the variation to ${\cal F}$, one obtains
\begin{equation}
d{\cal F} =  \sum_{pq>0} \tr
 \left(  \begin{array}{cc}  e_{pq}  &  -\Delta_{pq}  \\
             -\Delta_{qp}  &   -\overline{e}_{pq} \end{array} \right)
 \left(  \begin{array}{cc}  d\rho_{qp}  &  -d\kappa_{qp}  \\
             -d\kappa_{pq}  &  -d\overline{\rho}_{qp} \end{array} \right).
\end{equation}
with
\begin{equation}\label{4.11}
\displaystyle
e_{pq} =  \frac{1}{2} (1 +\delta_{pq})
\frac{\partial{\cal F}}{\partial\rho_{qp}}
,
\overline{e}_{pq} = \frac{1}{2}(1 +\delta_{pq}) \frac{\partial{\cal F}}
{\partial\overline{\rho}_{qp}}
,
\end{equation}
and
\begin{equation}
\Delta_{pq} = \frac{1}{2} \frac{\partial{\cal F} }{\partial\kappa_{pq}}.
\end{equation}
Defining
\begin{equation}\label{4.12}
\displaystyle
\mathcal H[\mathcal R] =  \left(  \begin{array}{cc}  e  &  -\Delta  \\
       -\Delta^{T}  &   -\overline{e} \end{array} \right),
\end{equation}
one deduces:
\begin{equation}
d {\cal G}[\mathcal R] =  \tr \left(\mathcal H[\mathcal R] - \Lambda \mathcal R - \mathcal R \Lambda +
             \Lambda  \right) \;  d\mathcal R.
\end{equation}
The stationary condition provides:
\begin{equation}
 \mathcal H[\mathcal R] - \Lambda \mathcal R - \mathcal R \Lambda +  \Lambda = 0.
\end{equation}
This leads to the Bogoliubov's equations, by elimination of $\Lambda$ with the help of Eq.(\ref{3.8}):
\begin{equation}
[ \mathcal H[\mathcal R], \mathcal R]=0.
\end{equation}

 \subsection{Gogny forces}


The Gogny forces have a long history in structure, reactions and fission studies, using microscopic
meanfield and beyond many-body approaches \cite{applications}. By essence, it's a two-body phenomenological effective
interaction which integrates a density-dependent term. In its original formulation, the central and
density non-dependent term was designed with finite ranges in order to include pairing correlations
at the HFB level, thus avoiding mainly ultraviolet divergences.

In the present first version of the HFB3 program, two families of Gogny forces have been implemented, a
family being defined here as a set of parameterizations based on the same analytical form for each type
of the term.

\noindent The {\it D1-type family} integrates parameterizations whose generic analytical form is written as
\begin{equation} \label{gognyD1}
\begin{split}
  \hat{v}_{12}^{\textrm{D1}} & = \sum_{i=1}^{N_c} G_{i}^{(c)} \hat{A}_{i}^{(c)} \\
  &  + t_0 (1+ x_0 \hat{P}_\sigma) \delta(\vec{r}_1 - \vec{r}_2) \rho^{\alpha} \Big( \frac{\vec{r}_1 + \vec{r}_2}{2} \Big) \\
&  + i W_0 \big[ \vec{k}' \times \delta(\vec{r}_1 - \vec{r}_2) \vec{k} \big] \cdot (\vec{\sigma}_1 + \vec{\sigma}_2)
\end{split}
\end{equation}
with
\begin{equation}
  G_{i}^{(x)} \equiv e^{-\frac{\left(\vec{r}_1-\vec{r}_2\right)^2}{\left(\mu_{i}^{(x)}\right)^2}}
\end{equation}
and
\begin{equation}
  \hat{A}_{i}^{(x)} \equiv W_{i}^{(x)} + B_{i}^{(x)} \hat{P}_\sigma - H_{i}^{(x)} \hat{P}_\tau - M_{i}^{(x)} \hat{P}_\sigma \hat{P}_\tau.
\end{equation}
One finds, by order of appearance, non-density-dependent Gaussian central terms, a zero-range density-dependent central term whose density is evaluated at the center of mass position and a zero-range spin-orbit
term. The HFB3 program contains the parameterizations D1 \cite{GognyD1}, D1S \cite{GognyD1S}, D1N \cite{GognyD1N}, D1M \cite{GognyD1M} for
which $N_{c}=2$ and D3G3 which has been built with an additional non-density-dependent Gaussian
central term such that $N_{c}=3$ \cite{GognyD3G3}.

\noindent The {\it D2-type family} extends the analytical form of the D1 family (\ref{gognyD1}) by generalizing the density-dependent term to a finite range term in such a way that
\begin{equation} \label{gognyD2}
\begin{split}
  \hat{v}_{12}^{\textrm{D2}} & \equiv \sum_{{i}=1}^{N_c} G_{i}^{(c)} \hat{A}_{i}^{(c)} \\
  &  + \sum_{{j}=1}^{N_d} \frac{1}{(\mu_{j}^{(d)} \sqrt{\pi})^3} \frac{\rho^{\alpha}(\vec{r}_1) + \rho^{\alpha}(\vec{r}_2)}{2} G_{j}^{(d)} \hat{A}_{j}^{(d)}  \\
& + i W_0 \big[ \vec{k}' \times \delta(\vec{r}_1 - \vec{r}_2) \vec{k} \big] \cdot (\vec{\sigma}_1 + \vec{\sigma}_2).
\end{split}
\end{equation}
In Eq. (\ref{gognyD2}), the density-dependent term has a Gaussian form factor. In the D2 parameterization \cite{GognyD2}, $N_{c}=2$ and $N_{d}=1$.

\noindent A future version of the HFB3 program incorporating {\it a fully finite range Gogny force} based on the D2 family but including finite range spin-orbit and tensor terms (the {\it DG-type family} \cite{GognyDG,Zietek25}) will be available in the future.

Finally, the Coulomb contribution to the total energy is evaluated either
exactly (in both mean and pairing fields) or with the help of the Slater approximation. By convention, the name of the interaction is completed by
a capital X,  when the Coulomb interaction is introduced without
approximation in the HFB solution. Thus, for example, D1S becomes D1SX,
D2 becomes D2X, etc\ldots

\section{Implementation specifics}
 \subsection{Axial symmetry}

 This version of the HFB3 program enforces the conservation of the axial symmetry along the $z$-axis. While this prevents the exploration of non-axial degrees of freedom, such as the triaxial nuclear shapes, which are known to lower the first fission barrier for several nuclear systems~\cite{Delaroche06}, this allows the formalism to make use of the associated $\Omega$-symmetry, resulting in an important numerical gain in the calculation of the interaction fields and of the nuclear observables.

Future versions of the HFB3 program may break this symmetry conservation and allow the exploration of non-axial nuclear shapes.

The geometrical symmetry $\mathbf{r}\rightarrow -\mathbf{r}$, also called nuclear parity, is explicitly broken by the HFB3 solver. Odd axial multipole moments, such as $\hat{Q}_{30}$ or $\hat{Q}_{50}$ for example, can then be constrained or optimized to non-zero values. In those cases, such multipole moments play a crucial role for the description of the nuclear fission process, since they can correspond to asymmetric fragmentations.

 \subsection{2-center HO basis}

In the HFB3 program, most matrix elements are evaluated in a 1-center or a 2-center basis. The 1-center basis corresponds to the widely used axial HO basis which can be found for example in the program HFBTHO \cite{Tho22}. The 2-center basis is the union of two identical 1-center bases shifted on the $z$-axis by $-\frac{d_0}{2}$ and $\frac{d_0}{2}$ respectively. This 2-center basis is well suited for the description of highly elongated shapes, but is also quite efficient for small to moderate elongations.

\subsubsection{Basis states\label{section_basis_states}}

Each state of the 1-center or 2-center bases is an HO solution uniquely determined by its quantum numbers $(m,n,n_z,d,s)$, all positive integers:
\begin{equation}
  | mnn_zds \rangle \equiv
  | mnn_zd \rangle \otimes
  | s \rangle.
\end{equation}
The wave function for the spatial part of the basis states in cylindrical coordinates $\mathbf{r}\equiv(r_\perp, \theta, z)$ is defined as
\begin{equation}
  \langle \mathbf{r} | mnn_zd \rangle
        \equiv
    Z(\zeta, n_z)
    .
    R(\eta, m, n)
    .
         e^{im\theta}
\end{equation}
with
\begin{eqnarray}
Z(\zeta, n_z)
  &\equiv&
N_{n_z}^{b_z}
         e^{-\frac{\zeta^2}{2}}H_{n_z}(\zeta)  \\
R(\eta, m, n)
         &\equiv&
         N_{m,n}^{b_\perp}
         e^{-\frac{\eta}{2}}
         \eta^{\frac{|m|}{2}}
         L_n^{|m|}\left(\eta\right),
\end{eqnarray}
using the transformed coordinates $\eta$ and $\zeta$ defined as
\begin{equation}
  \eta \equiv \left(\frac{r_\perp}{b_\perp}\right)^2,
  \zeta \equiv \frac{z-d_0 \left(\frac{1}{2} - d\right)}{b_z}.
\end{equation}
The three basis parameters are $(b_\perp, b_z, d_0)$.
The normalization constants are defined as
\begin{equation}
  N_{n_z}^{b_z}
  \equiv
         \frac{1}{\sqrt{b_z}}
         \frac{1}{\sqrt{2^{n_z} \sqrt{\pi}n_z!}},
\end{equation}
and
\begin{equation}
  N_{m,n}^{b_\perp}
  \equiv
  \frac{1}{b_{\perp}\sqrt{\pi}}
  \sqrt{\frac{n!}{(n+|m|)!}}.
\end{equation}
The $H_{n_z}$ and $L_n^{|m|}$ functions are the usual Hermite and Generalized Laguerre polynomials, respectively. The spin part of each basis state is determined by the $s$ quantum number (spin up and spin down for values 0 and 1, respectively).
For a generic 2-center basis, $d\in\{0,1\}$ and $d_0>0$. The special case of a 1-center basis is obtained by setting $d=0$ and $d_0=0$.

\subsubsection{Basis orthogonalization}

The implementation of the HFB formalism in this 2-center HO basis differs from the 1-center HO case mainly in the computation and use of special transformation matrices between the non-orthogonal 2-center basis and a slightly smaller bi-orthogonal basis. The associated representations are noted HO and OR, respectively. In the HFB3 implementation, most of the representation-dependent quantities have been calculated in the HO representation (multipole moments, interaction fields), some in the OR representation (HFB hamiltonian matrix), and some in the spatial  $(r_\perp, z)$ representation (fragment properties, radii).
The procedure to construct and use the OR basis has been described in great details in \cite{bergerPhd}.

\subsubsection{Basis truncation}

The set of quantum numbers included in the HO basis is given by the following truncation procedure involving the basis truncation parameters $N$ and $Q$. We define
\begin{equation}
  \forall i \in \mathbb{N}^*, \nu(i) \equiv (N+2).Q^{\frac{2}{3}} + \frac{1}{2} - i.Q, \label{defQ}
\end{equation}
and we find
\begin{equation}
m^{\textrm{max}} \equiv \textrm{sup}\{i:\nu(i)\ge 1\}.
\end{equation}
The set of quantum numbers $(m, n, n_z)$ is then defined as
\begin{eqnarray}
  0 & \le m   < & m^{\textrm{max}},\\
  0 & \le n   < & \frac{1}{2}\left(m^{\textrm{max}}-m-1\right)+1,\\
  0 & \le n_z < & \nu(m+2n+1).
\end{eqnarray}

The basis truncation parameters can have a significant impact on the resulting state.
In particular, for the description of very elongated nuclear systems, as is the case for the description of nuclear fission, 2-center bases are better suited than size-equivalent 1-center bases, as can be seen in section \ref{basisSize}.

\subsection{Minimization methods}


The search for the CHFB minimum can be performed following two alternative methods. Both are in essence equivalent (converged HFB vacua are expected to be identical modulo a unitary gauge transformation) but present complementary advantages. Self-consistent iterations determine \(\mathcal R\) matrices via a fixed point method, while gradient descent minimizes \(\mathcal E(\mathcal R)\) under a set of constraints.
Both methods are implemented in HFB3 in the classes \texttt{SolverHFBBroyden} and \texttt{SolverHFBGradient}, respectively.

\subsubsection{Broyden mixing}

Given a density matrix \(\mathcal R_k\) at iteration \(k\), one self-consistent HFB iteration corresponds to one update of \(\mathcal R_k\) such that the new matrix \(\mathcal R_{k+1}=f(\mathcal R_{k})\) commutes with \(\mathcal H[\mathcal R_k]\). Fixed points of this iteration function naturally verify \([\mathcal R, \mathcal H[\mathcal R]]=0\). However, fixed point methods often exhibit slow or even non converging behaviors.

Broyden method \cite{Bro65} helps with convergence by constructing \(\mathcal R_{k+1}\) as a mixing between \(f(\mathcal R_k)\) and all previous iterations \(\{\mathcal R_0\, \cdots,\mathcal R_k\}\) used to construct an approximation to the inverse Jacobi of \(f\). In order to avoid numerical instabilities and speed up the calculations, the inverse Jacobi is approximated by using solely a subset of the previous iterations.

 \subsubsection{Gradient descent}

 Formulation of gradient method minimizes \(E[\mathcal R]\) by unitarily evolving \(\mathcal R\) along the steepest descent. The analytical expressions of the gradient, not given here, are based on the use of Thouless theorem to parameterize the set of neighboring Bogoliubov states. Lagrange parameters are naturally added to the formalism to handle constraints of any multipole operator. In order to improve convergence properties, gradient with momentum \cite{Rob11} is implemented.

\subsubsection{\label{alternatingSolvers}Alternating solvers}

It is possible to alternate between the \texttt{SolverHFBBroyden} and \texttt{SolverHFBGradient} classes at any point during the convergence process. Most of the time, this is done either to initiate a convergence 'from scratch' with the Gradient method (rock-solid convergence even without an initial neighboring solution), or to speed-up a well-behaving ongoing convergence with the Broyden method (very fast convergence if the solution is relatively close). If no initial solution is given, one can also initiate the convergence with a Woods-Saxon solution. The class \texttt{SolverWS} implements a very basic deformed Woods-Saxon solver (no pairing, spin-orbit or Coulomb) with an automatic deformation parameters optimization to match the HFB deformation constraint values. This solver is much faster than the Gradient one, but in some rare cases, its solution may lead to convergence issues in the Broyden solver. One could then try to start the same calculation with \texttt{SolverHFBGradient} instead.

 \subsection{Constraints}

It is possible to use an arbitrary number of constraints during HFB convergence. Such constraints can impose a specific mean-value of the usual multipole moment operators or some 1-body geometry properties (neck operator, fragment mass asymmetry, etc\ldots).

\subsubsection{Multipole moment operators}

Since other HFB solvers may use different normalization factors for the intrinsic mass multipole moments, one may have to convert mass multipole mean values to be able to compare results between HFB solvers.
The program HFB3 uses the axial intrinsic mass multipole moments defined as (cf. (B-18) in \cite{ring})
\begin{equation}
\hat{Q}_{\lambda 0} \equiv r^\lambda \hat{Y}_{\lambda 0},
\end{equation}
with $\hat{Y}_{\lambda 0}$ being the usual spherical harmonics. For reference, we give the expressions of the first mass multipole moments:
\begin{align}
\label{q00def} \hat{Q}_{00} & = \frac{1}{ 2}\sqrt{\frac{ 1}{\pi}}, \\
\label{q10def} \hat{Q}_{10} & = \frac{1}{ 2}\sqrt{\frac{ 3}{\pi}} \left(z                                                        \right), \\
\label{q20def} \hat{Q}_{20} & = \frac{1}{ 4}\sqrt{\frac{ 5}{\pi}} \left(2z^2 - r_\perp^2                                         \right), \\
\label{q30def} \hat{Q}_{30} & = \frac{1}{ 4}\sqrt{\frac{ 7}{\pi}} \left(2z^3 - 3z r_\perp^2                                      \right), \\
\label{q40def} \hat{Q}_{40} & = \frac{3}{16}\sqrt{\frac{ 1}{\pi}} \left(8z^4 - 24z^2r_\perp^2 +3 r_\perp^4                       \right), \\
\label{q50def} \hat{Q}_{50} & = \frac{1}{16}\sqrt{\frac{11}{\pi}} \left(8z^5 - 40z^3r_\perp^2 + 15 z r_\perp^4                   \right), \\
\label{q60def} \hat{Q}_{60} & = \frac{1}{32}\sqrt{\frac{13}{\pi}} \left(16z^6 - 120z^4r_\perp^2 + 90 z^2 r_\perp^4 -5 r_\perp^6  \right).
\end{align}
These expressions use the cylindrical coordinates $z$ and $r_\perp$ defined in \ref{section_basis_states}.
The program HFB3 can evaluate the mean value of these multipole moments analytically or numerically using a Gauss-type quadrature integration technique\cite{Gradshteyn65}.

 Some constraints can be placed on the mean value of some of these multipole moment operators.
 Since the unconstrained HFB hamiltonian is invariant under any $z$-translation, a constraint on the mean value of the mass $\hat{Q}_{10}$ operator is expected for any HFB calculation (the program HFB3 will issue a warning if this is not the case).
 Any combination of additional constraints on the mean value of the first multipole moments is allowed with $\lambda \le 6$.
 Constraints can be placed on the mean value of mass, proton and/or neutron multipole moments, however using simultaneous constraints on protons, neutrons and total (protons and neutrons) for a given multipole moment should be avoided, since this may lead to convergence issues.

 For convenience, it is also possible to set constraints on the usual $\hat{\beta}_{20}$ elongation operator, defined as
\begin{equation}
  \hat{\beta}_{20} \equiv \frac{4\pi}{3 r_0^2 A^{5/3}}\hat{Q}_{20}
\end{equation}
with $r_0=1.2$ fm. If such a constraint is specified, it is automatically converted to a constraint on $\hat{Q}_{20}$.

The HFB3 program can be instructed to use the multipole moment definitions and physical constant values from another solver, through the use of `compatibility modes'. This can be done by using the following key/value pair in the input \texttt{DataTree}\label{compatModes}:
\begin{verbatim}
key: general/compatibility
val: {'', robledo, berger, htbtho}
\end{verbatim}
The possible values are associated to the solvers written by L.~Robledo, J.-F.~Berger and the HFBTHO team, respectively.

 \subsubsection{Lagrange multipliers adjustment}

 The constraints are implemented using Lagrange multipliers. For a given isospin $\nu$ and an associated set of constraints $\mathcal{C}^\nu$, the constrained HFB hamiltonian reads
 \begin{equation}
   \hat{h}_{\textrm{CHFB}}^{\nu} \equiv  \hat{h}_{\textrm{HFB}}^{\nu} - \sum_{j\in \mathcal{C}^\nu} \lambda_j^{\nu} \hat{Q}_{j0}^{\nu}
 \end{equation}
 with $\lambda_{j}^{\nu}$ being the Lagrange multipliers which are adjusted during the HFB iterations of the solver. This adjustment uses an inversion of the QRPA matrix, which is done in an approximate way by using a cranking technique (cf. p.163 in \cite{bergerPhd} for details).
 The constrained values are set by imposing the following additional conditions on the final state $|\psi\rangle$
 \begin{equation}
   \forall j\in \mathcal{C}^\nu,   \langle \psi^\nu | \hat{Q}_{j0}^\nu | \psi^\nu \rangle = q_{j0}^\nu.
 \end{equation}

\subsection{Collective inertia tensors}

The collective inertia tensors and their associated quantities can be calculated by the HFB3 program using two different approximations: the Adiabatic Time-Dependent Hartree-Fock-Bogoliubov (ATDHFB) or the Generator Coordinate Method (GCM). Full descriptions of these approximations can be found in \cite{ring} and \cite{baranger}, we only give the final expressions here. The metric $\Gamma$, the zero-point energies $\epsilon_{\rm ZPE}$ and the collective inertia tensors $\mathrm{B}$ can be expressed in terms of the moments $\mathrm{M}^{(k)}_{ij}$ by using the cranking approximation for the QRPA matrix:

\begin{align}
\Gamma &= \frac{1}{2}\left(\mathrm{M}^{(1)}\right)^{-1}\mathrm{M}^{(2)}\left(\mathrm{M}^{(1)}\right)^{-1},\\
\mathrm{B}^{\rm GCM} &= \left[ 4 \Gamma \mathrm{M}^{(1)}\Gamma\right]^{-1},\\
\mathrm{B}^{\rm ATDHFB} &= \left[ \left( \mathrm{M}^{(1)}\right)^{-1} \mathrm{M}^{(3)} \left( \mathrm{M}^{(1)}\right)^{-1}  \right]^{-1},\\
\epsilon_{\rm ZPE}^{\rm GCM/ATDHFB} &= \frac{1}{2} \Gamma \mathrm{B}^{\rm GCM/ATDHFB}
\end{align}
with
\begin{equation}
\mathrm{M}^{(k)}_{ij} = \sum_{l<m} \frac{\langle\psi|\hat{Q}^\dagger_{i} \beta^\dagger_l \beta^\dagger_m| \psi \rangle \langle \psi|\beta_m\beta_l \hat{Q}^\dagger_{j}  |\psi\rangle}{(E_l + E_m)^k},
\end{equation}
where $\beta^\dagger_x$ is the quasiparticle $x$ creation operator and $E_x$ is the energy associated with the quasiparticle $x$.

\subsection{Odd systems}

In order to describe even-odd, odd-even or odd-odd systems, a blocking technique using the Equal Filling Approximation (EFA) is used in HFB3 (cf. for example \cite{PerezMartin08}). If such an odd system is considered, the program will first calculate the closest lighter even-even system with the same constraints, and will perform several successive HFB calculations with different blocked quasi-particle states. The blocked quasi-particle states are chosen using the quasi-particle spectrum of the even-even system. Each blocking trial will start from the even-even HFB state, to speed up the HFB loop. The final state will then be chosen among the blocking trials as the one minimizing the total binding energy. If an HFB solution of the closest lighter even-even system is given as a starting point, its quasi-particle spectrum is used directly. For performance reasons, it is advised to optimize the basis parameters for the even-even system, and to then keep these basis parameters when performing the blocking trials. This allows for a rather fast HFB convergence in an almost-optimal basis.

An example of such chained even-even and even-odd calculations is given in files

\begin{itemize}
  \item \texttt{examples/90Zr\_deformed\_1x9.hfb3} (even-even),
  \item \texttt{examples/91Zr\_deformed\_1x9.hfb3} (odd-even).
\end{itemize}
By default, the number of blocking trials for each isospin value is set to 4, but this value can be changed with the key:value pair `\texttt{action/nbBlockingTrials}' in the input \texttt{DataTree}.

\section{Using the program}

The instructions to install and use the HFB3 program are given in the README.md file. We will summarize them here.
There are two ways to use the HFB3 program: either by using a command-line interface (CLI) executable or by using the Python bindings.
Both ways give access to equivalent functionalities and should have equivalent performances.

 \subsection{Installation}

There are two ways to install the HFB3 program: from source or from the Python Package Index (PyPI).

 \subsubsection{From source}

One can build and install the HFB3 CLI executable and the HFB3 Python bindings from source by getting a release from its official repository and deploying it.
If the requirements are fulfilled, issuing the command `\texttt{make}' at the root of the project will compile and link the HFB3 CLI executable in `\texttt{bin/hfb3}'. The command
\begin{verbatim}
$ make python_install
\end{verbatim}
will generate the Python bindings and install them in the current Python environment. The command
\begin{verbatim}
$ make doc
\end{verbatim}
will use \texttt{Doxygen} \cite{doxygen} to build the technical documentation. The main page for the HTML documentation can be found in `\texttt{doc/html/index/html}'.

 \subsubsection{From PyPI}

If one is only interested in the Python bindings of HFB3, one can get and install them directly from PyPI\cite{PyPI} by using the command
\begin{verbatim}
$ pip install hfb3
\end{verbatim}
This will install the bindings and some examples, but not the main HFB3 CLI executable, nor the documentation.

\subsection{Example runs (CLI)}

The HFB3 CLI executable takes several command-line options. Use `\texttt{hfb3 -h}' to list them. In particular, the option `\texttt{--list-keys}' prints a list of the keys that can be used in a \texttt{DataTree} instance, their meaning, type and default value.
Some calculation examples are given in the `\texttt{examples/}' directory. To launch a calculation, the HFB3 program first has to construct a \texttt{DataTree} instance, which can be seen as an input dictionary. Every CLI argument given to the HFB3 CLI executable will be interpreted first as a possible CLI option, then as a filename, then as a \texttt{JSON}-like content. The HFB3 program can parse several file formats to construct a \texttt{DataTree} instance: MessagePack (like a binary \texttt{JSON})~\cite{msgp}, \texttt{.hfb3} (similar to a \texttt{.ini} file), and \texttt{JSON}-like content.

To calculate the lowest-energy axial HFB state of $^{16}$O under the simultaneous constraints $\langle\hat{Q}_{10}\rangle = 0\textrm{fm}$, $\langle\hat{Q}_{20}\rangle = 8\textrm{fm}^2$ and $\langle\hat{Q}_{30}\rangle = 0\textrm{fm}^3$, in an optimized 7-shells 1-center basis, using the D1S interaction, one can simply use the command line
\begin{verbatim}
$ bin/hfb3 examples/16O_deformed.hfb3
\end{verbatim}
The total 1-body local density can be plotted using the following Python script (result shown in Fig. \ref{16ODensity}):
\begin{verbatim}
$ bin/plotLocalDensity.py
\end{verbatim}

\begin{figure}[ht]
  \centering
  \includegraphics[width=1.00\columnwidth]{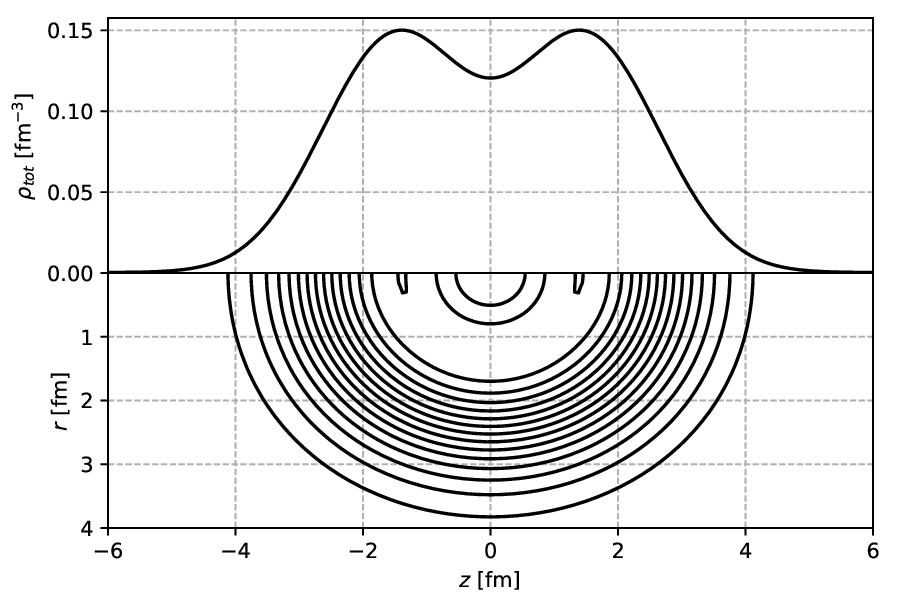}
  \caption{\label{16ODensity}Total 1-body local density for a deformed solution of $^{16}$O. In the top panel, the curve represents the axial density ($r=0$ fm). The isolines are separated by $0.01$ fm$^{-3}$.}
\end{figure}

\begin{figure}[ht]
  \centering
  \includegraphics[width=1.00\columnwidth]{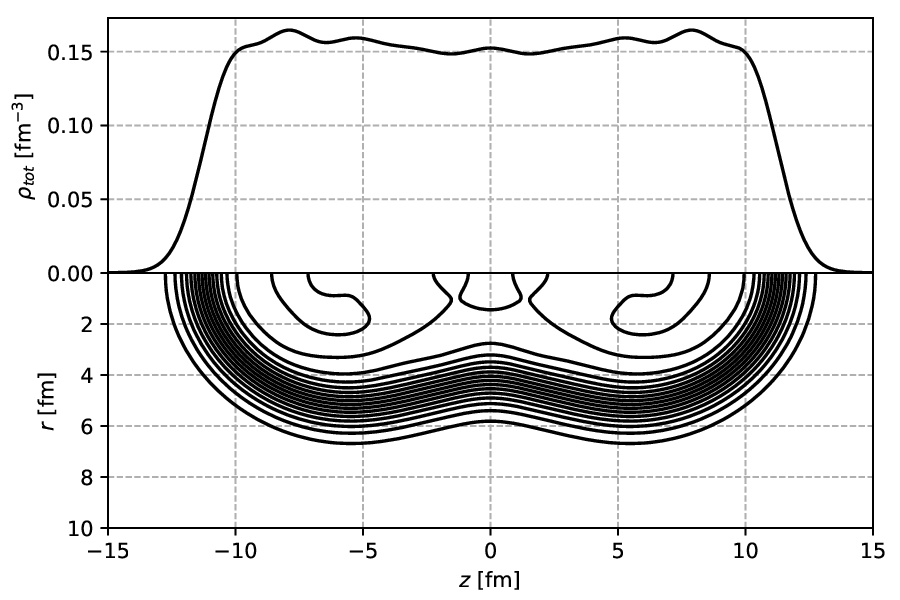}
  \caption{\label{240PuDensity}Total 1-body local density for a deformed solution of $^{240}$Pu. In the top panel, the curve represents the axial density ($r=0$ fm). The isolines are separated by $0.01$ fm$^{-3}$.}
\end{figure}

One can perform a similar calculation for the $^{240}$Pu nuclear system by using the command line
\begin{verbatim}
$ bin/hfb3 examples/240Pu_deformed.hfb3
\end{verbatim}
For this example calculation, the constraints are $\langle\hat{Q}_{10}\rangle = 0\textrm{ fm}$ and $\langle\hat{\beta}_{20}\rangle = 1.4$. The basis used is an optimized 2-center, 11 major HO shells one (can also be called a `2x11 basis'). Using the previously mentioned Python script, one can generate Fig. \ref{240PuDensity}.

\subsection{Example notebook (Python)}

To illustrate how HFB3 can be used from Python, a Jupyter notebook showing several manipulations of the main classes is proposed. It can be used with the command
\begin{verbatim}
$ jupyter notebook examples/demo.ipynb
\end{verbatim}

Depending on the Python environment of the user, some Python dependencies may be needed in order to execute this notebook.
A simple Python script is proposed and can be used to replicate the previous example for the calculation of a deformed $^{16}$O HFB state with the command
\begin{verbatim}
$ python3 examples/16O_deformed.py
\end{verbatim}

\subsection{Potential Energy Surfaces}

To generate 1-dimensional potential energy surfaces (1D-PES) using multiple processes, a Python script is provided in `\texttt{misc/studies/pes\_generator}'.
The main deformation constraint is $\langle \hat{\beta}_{20}\rangle$, but this can be easily changed by tweaking the script.
A retro-propagation algorithm is implemented to ensure that detected local minima are automatically replaced by global minima.
The basis deformation parameters are optimized for each calculation.
We have used this script to generate the various 1D-PESs in the following part.

\subsection{Basis types and sizes\label{basisSize}}

To check the influence of the basis truncation parameters on the resulting constrained HFB states, several 1D-PESs have been produced for $^{240}$Pu with the Gogny D1S interaction, for different basis types (1-center and 2-center) and basis sizes. Each curve consists of 601 converged HFB solutions, allowing spontaneous breaking of the parity symmetry ($z\rightarrow -z$). It is worth noting that the generation of these solutions has been done without any human intervention nor cherry-picking of the results, showing the robustness of the solvers.

\begin{figure}[ht]
  \centering
  \includegraphics[width=0.95\columnwidth]{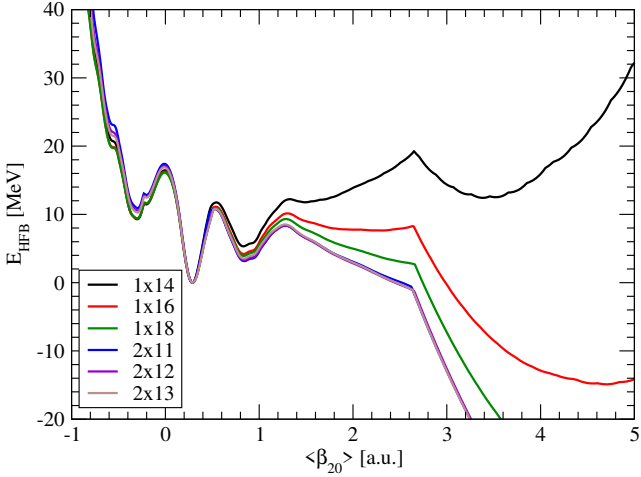}
  \caption{\label{pesRel}Potential energy surfaces for $^{240}$Pu with different basis types and sizes. The label '1x14' indicates a 1-center basis using 14 major harmonic oscillator shells. All bases are using $Q=1.0$. The total binding energy is relative to the normal deformed minimum ($\langle \hat{\beta}_{20}\rangle=0.29$). The oscillations visible on the highly deformed part of the '1x14' curve are due to the quantization of the basis deformation parameters.}
\end{figure}

\begin{figure}[ht]
\centering
  \includegraphics[width=0.95\columnwidth]{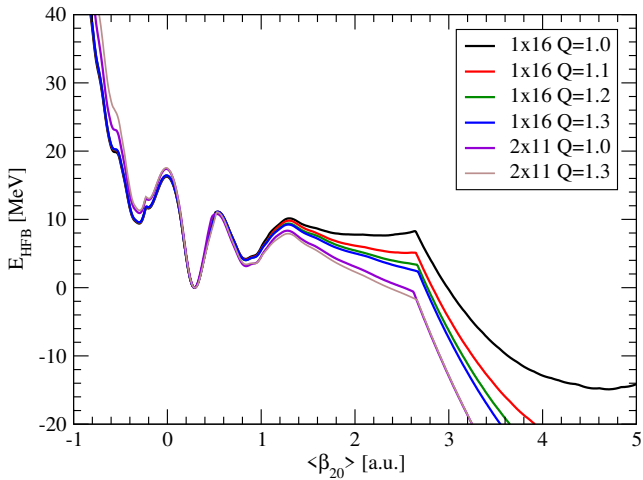}
  \caption{\label{pesQ}Comparison of potential energy surfaces for different $Q$ values. The 2-center bases are less sensitive than 1-center ones to changes of this basis truncation parameter.}
\end{figure}

One can see on Fig. \ref{pesRel} that using 1-center bases induces a spurious energy drift for highly elongated systems. This drift can be reduced by increasing the basis size, at the cost of memory usage and calculation length (as can be seen in Table \ref{sizeAndMem}). One can also increase the value of $Q$ (Eq. \eqref{defQ}), as can be seen on Fig. \ref{pesQ}. However, even with higher $Q$ values, 1-center bases always present a non-negligible spurious energy drift compared to total binding energies obtained with equivalent 2-center bases at high elongations. Using such 2-center bases avoids this spurious energy drift, even for relatively modest basis sizes.

\begin{figure}[ht]
  \centering
\includegraphics[width=0.95\columnwidth]{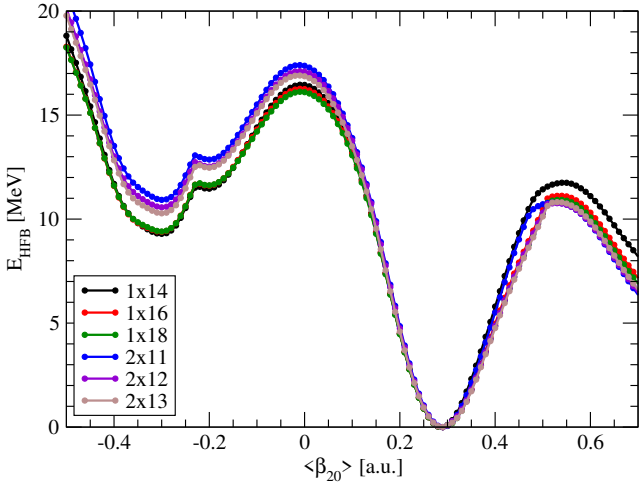}
  \caption{\label{pesRelZoom}Zoom on the low-deformed part of Fig. \ref{pesRel}. Each point represents a converged HFB solution.}
\end{figure}

For big 1-center bases with high values of $Q$ (typically $N>16$ and $Q>1.3$), some HO states may correspond to high $n_z$ values (higher that 24). Including such states in the basis may lead to numerical inaccuracies, due to the limited precision of the encoding of double-precision floating point numbers (using 64 bits). This is why such states are not included in the basis. The default criterion for this rejection ($n_z > 24$) can be changed with the key `\texttt{basis/n\_zMax}' in the input \texttt{DataTree}. If an HFB state presents significant contributions from high-valued $n_z$ HO states, such missing states can lead to issues with the HFB solver, mainly characterized by a highly non-physical total binding energy and/or multipole moments.
It is strongly advised to avoid the use of such highly truncated big 1-center bases with the current version of HFB3. In case of issues, consider using a lower $Q$ value or a 2-center basis (cf. the following discussion about the evolution of 2-center bases in Figs. \ref{pesRel} and \ref{pesRelZoom}).

\begin{figure}[ht]
\centering
\includegraphics[width=0.95\columnwidth]{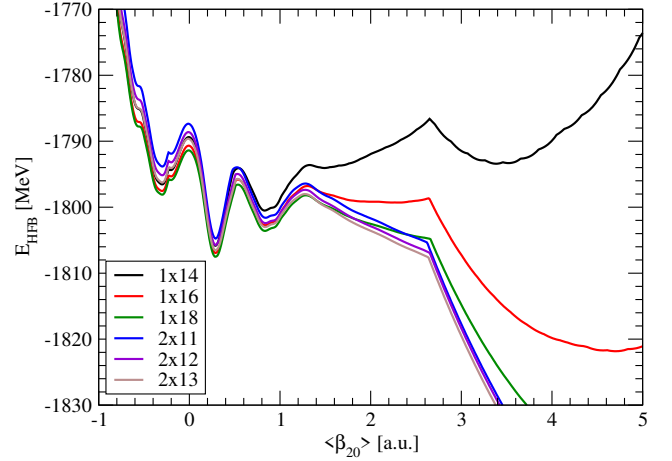}
  \caption{\label{pesAbs}Same as Fig. \ref{pesRel}, but the total binding energy is absolute.}
\end{figure}

\begin{figure}[ht]
\centering
  \includegraphics[width=0.95\columnwidth]{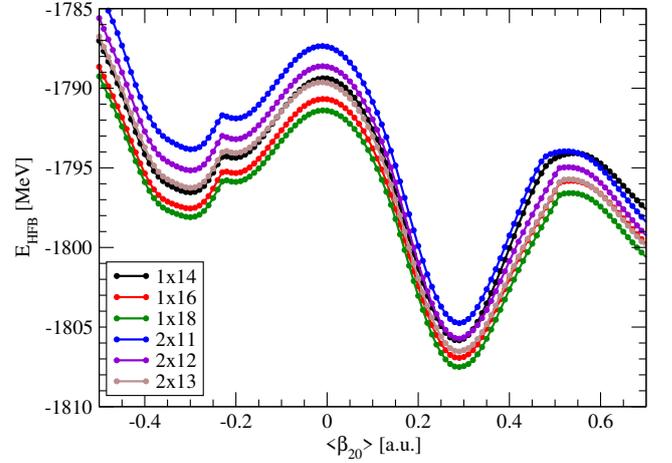}
  \caption{\label{pesAbsZoom}Zoom on the low-deformed part of Fig. \ref{pesAbs}. Each point represents a converged HFB solution.}
\end{figure}

As can be seen on Fig. \ref{pesRelZoom}, using 2-center bases can however induce a moderate spurious energy drift for oblate solutions, compared to size-equivalent 1-center calculations.
In Figs. \ref{pesAbs} and \ref{pesAbsZoom}, the absolute total binding energy is represented for the same bases as in Figs. \ref{pesRel} and \ref{pesRelZoom}. One can see that the curves corresponding to 1-center and 2-center bases are globaly shifted when increasing the basis size. In cases where the absolute total binding matters, for example if one wishes to study nuclear masses, the bigger the basis, the better. However, some techniques can be used in order to avoid too costly calculations, such as the `infinite basis correction' (cf. Eq. (2) and discussion in \cite{Hilaire07}).

For a given basis, it is expected that the smaller the nuclear system, the smaller the energy drifts with the deformation. This study of $^{240}$Pu can then be considered as a worst-case scenario to study these drifts.
Depending on the range of deformations used in a study and the mass of the nuclear system, some bases presenting the best compromises between execution length, memory usage and numerical precision can be highlighted. In Table \ref{sizeAndMem}, we have indicated both of these `recommended bases' for structure-only and structure+fission studies of actinides. In particular, the `2x11' basis presents a moderate energy drift for oblate shapes and a well converged description of the elongated shapes, for a rather cheap numerical cost (less than 1s per iteration, around 1.3 GB of memory usage, and around 2MB per result file).

\begin{table}[ht]
  \begin{tabular}{rrrrr}
   \toprule
    \cell{Basis\\ type} & \cell{Number\\ of states} & \cell{Iteration\\ length [s]} & \cell{Memory\\ usage [MB]}    & \cell{Result file\\ size [MB]} \\
   \hline
    1x13 & 560 & 0.354 &  663 &  1.267 \\
   1x14 &  680 & 0.528 &  925 &  1.771 \\
   1x15 &  816 & 0.858 & 1315 &  2.409 \\
  *1x16 &  969 & 1.070 & 1842 &  3.209 \\
   1x17 & 1140 & 1.490 & 2394 &  4.240 \\
   1x18 & 1330 & 2.130 & 3263 &  5.484 \\
   1x19 & 1540 & 2.920 & 4323 &  7.015 \\
   1x20 & 1771 & 4.780 & 5752 &  8.838 \\
   2x09 &  440 & 0.329 &  473 &  0.861 \\
   2x10 &  572 & 0.592 &  783 &  1.373 \\
  *2x11 &  728 & 0.871 & 1229 &  2.097 \\
   2x12 &  910 & 1.370 & 1905 &  3.065 \\
   2x13 & 1120 & 2.020 & 2766 &  4.578 \\
   2x14 & 1360 & 3.290 & 4147 &  6.332 \\
   2x15 & 1632 & 4.530 & 5859 &  8.938 \\
   2x16 & 1938 & 7.140 & 8345 & 12.043 \\
 \end{tabular}
  \caption{\label{sizeAndMem}Comparison of number of states, average iteration length, memory used and result file size (compressed with gzip) for different basis types and sizes. The CPU used is a 3.8GHz Intel Core Ultra 9 185H. Each calculation is using only 1 CPU core. The starred lines correspond to recommended bases for structure-only and structure+fission studies of actinides, respectively.}
\end{table}

\section{Benchmark with HFBTHO}

In the following, some constrained HFB calculations will be performed with HFB3 and HFBTHO, a publicly available axial HFB solver with zero- and finite-range interactions using a 1-center HO basis~\cite{HFBTHOv4}.
The goal of these calculations is to compare the results and the efficiency of both codes for almost-identical physical cases. The input and output files for both codes for each physical case can be found in `\texttt{misc/studies/benchmark\_hfbtho}'.

In order to use identical conventions and physical constants for both codes, the compatibility mode `\texttt{hfbtho}' has to be activated in HFB3 (cf. Section \ref{compatModes}). In this compatibility mode, the HFB3 solver is modified:

\begin{itemize}
  \item the 2-body-center-of-mass correction field~\cite{Bender,Butler} is disabled (HFBTHO does not include it for Gogny-type interactions~\cite{HFBTHOv3}),
  \item the mass factor in the kinetic field takes a slightly different value,
  \item the global factor in the Coulomb field (Slater approximation) takes a slightly different value,
  \item the first three multipole moments are defined as
\end{itemize}
\begin{align}
  \hat{Q}_{00} & \equiv 1, \\
  \hat{Q}_{10} & \equiv \hat{z},\\
  \hat{Q}_{20} & \equiv 2 \left(2\hat{z}^2 - \hat{r}_\perp^2 \right).
\end{align}

To achieve an identical behavior between both codes, a change must be made in HFBTHO: the rearrangement term coming from the density-dependence of the Slater approximation of the Coulomb Exchange field has to be disabled, since this field calculation is not implemented in HFB3 (HFB3 allows an exact calculation of the Coulomb field, so there is no plan to implement the missing rearrangement field).

The calculations have been performed on the same hardware (a 3.8GHz Intel Core Ultra 9 185H), using one process on one CPU core (enabling OpenMP multi-threading slows down both codes). All proposed compiler optimizations have been enabled.

We present in Tables~\ref{table:20O} and \ref{table:208Pb} the results obtained for two spherical systems $^{20}$O and $^{208}$Pb. The convergence at iteration $k$ noted $\epsilon_k$ is defined as a function of the density matrix $\rho$ in the HO representation for both isospins:
\begin{equation}
  \epsilon_k \equiv   || \rho^n_k - \rho^n_{k-1}||_{\infty}
                    + || \rho^p_k - \rho^p_{k-1}||_{\infty}.
\end{equation}

\begin{table}[ht]
  \centering
  \begin{tabular}{l|c|r|r}
   \toprule
    Quantity          &  Unit & HFB3               & HFBTHO             \\ \hline
    \#iterations      &       & $25$               & $24$               \\ \hline
    Runtime           & [s]   & $1.466$            & $32.114$           \\ \hline \hline
    HFB energy        & [MeV] & $-159.160$         & $-159.160$         \\ \hline
    Kinetic (dir.)    & [MeV] & $317.941$          & $317.941$          \\ \hline
    Central (dir.)    & [MeV] & $-871.601$         & $-871.601$         \\ \hline
    Central (exch.)   & [MeV] & $-63.71\textbf{1}$ & $-63.71\textbf{0}$ \\ \hline
    Central (pair.)   & [MeV] & $-6.421$           & $-6.421$           \\ \hline
    Coulomb (dir.)    & [MeV] & $16.333$           & $16.333$           \\ \hline
    Coulomb (exch.)   & [MeV] & $-2.812$           & $-2.812$           \\ \hline
    Spin-Orbit (dir.) & [MeV] & $-15.442$          & $-15.442$          \\ \hline
    Density (dir.)    & [MeV] & $466.55\textbf{1}$ & $466.55\textbf{2}$ \\ \hline
    RMS radius        & [fm]  & $2.795$            & $2.795$            \\ \hline
    Charge radius     & [fm]  & $2.794$            & $2.794$            \\
  \end{tabular}
  \caption{Comparison for an HFB calculation of $^{20}$O with the D1S Gogny functional. The `1x10' basis uses the deformation and truncation parameters $b_r = b_z = 1.457$fm and $Q = 1.0$. Mass multipole moments $\hat{Q}_{10}$ to $\hat{Q}_{60}$ have been constrained to zero. The convergence target is set to $\epsilon_{\textrm{max}}=10^{-8}$.}
  \label{table:20O}
\end{table}

\begin{table}[ht]
  \centering
  \begin{tabular}{l|c|r|r}
   \toprule
    Quantity          &  Unit & HFB3                  & HFBTHO                \\ \hline
    \#iterations      &       & $31$                  & $31$                  \\ \hline
    Runtime           & [s]   & $49.613$              & $379.130$             \\ \hline \hline
    HFB energy        & [MeV] & $-1650.750$           & $-1650.750$           \\ \hline
    Kinetic (dir.)    & [MeV] & $3914.14\textbf{6}$   & $3914.14\textbf{4}$   \\ \hline
    Central (dir.)    & [MeV] & $-12154.1\textbf{39}$ & $-12154.1\textbf{25}$ \\ \hline
    Central (exch.)   & [MeV] & $-616.72\textbf{1}$   & $-616.72\textbf{2}$   \\ \hline
    Coulomb (dir.)    & [MeV] & $830.63\textbf{5}$    & $830.63\textbf{4}$    \\ \hline
    Coulomb (exch.)   & [MeV] & $-31.377$             & $-31.377$             \\ \hline
    Spin-Orbit (dir.) & [MeV] & $-104.684$            & $-104.684$            \\ \hline
    Density (dir.)    & [MeV] & $6511.38\textbf{9}$   & $6511.38\textbf{0}$   \\ \hline
    RMS radius        & [fm]  & $5.520$               & $5.520$               \\ \hline
    Charge radius     & [fm]  & $5.497$               & $5.497$               \\
  \end{tabular}
  \caption{Comparison for an HFB calculation of $^{208}$Pb with the D1S Gogny functional. The `1x16' basis uses the deformation and truncation parameters $b_r = b_z = 2.223$fm and $Q = 1.0$. Mass multipole moments $\hat{Q}_{10}$ to $\hat{Q}_{60}$ have been constrained to zero. The convergence target is set to $\epsilon_{\textrm{max}}=10^{-8}$.}
  \label{table:208Pb}
\end{table}

One can see that the results are quite similar, the only deviations being related to numerical noise or order of operations.
Even if some field energy contributions show minor differences (a few keV), the total binding energy difference between both codes is lower than 1keV.
The RMS and charge radii are identical. Even if the number of HFB iterations is similar, the total execution time is clearly in favor of HFB3, which is between 8 and 20 times faster, depending on the system considered.

We present in Table~\ref{table:202Rn} the results obtained for the deformed system $^{202}$Rn.

\begin{table}[ht]
  \centering
  \begin{tabular}{l|c|r|r}
   \toprule
    Quantity          &  Unit       & HFB3                  & HFBTHO                \\ \hline
    \#iterations      &             & $37$                  & $54$                  \\ \hline
    Runtime           & [s]         & $62.960$              & $688.210$             \\ \hline \hline
    HFB energy        & [MeV]       & $-1570.863$           & $-1570.863$           \\ \hline
    Kinetic (dir.)    & [MeV]       & $3744.247$            & $3744.247$            \\ \hline
    Central (dir.)    & [MeV]       & $-11727.4\textbf{60}$ & $-11727.4\textbf{57}$ \\ \hline
    Central (exch.)   & [MeV]       & $-63.231$             & $-606.231$            \\ \hline
    Central (pair.)   & [MeV]       & $-25.262$             & $-25.262$             \\ \hline
    Coulomb (dir.)    & [MeV]       & $897.685$             & $897.685$             \\ \hline
    Coulomb (exch.)   & [MeV]       & $-33.230$             & $-33.230$             \\ \hline
    Spin-Orbit (dir.) & [MeV]       & $-83.882$             & $-83.882$             \\ \hline
    Density (dir.)    & [MeV]       & $6263.26\textbf{9}$   & $6263.26\textbf{7}$   \\ \hline
    $Q_{20}^{(n)}$    & [fm$^2$]    & $1702.458$            & $1702.458$            \\ \hline
    $Q_{20}^{(p)}$    & [fm$^2$]    & $1297.542$            & $1297.542$            \\ \hline
    $Q_{30}^{(n)}$    & [fm$^3$]    & $557.539$             & $557.539$             \\ \hline
    $Q_{30}^{(p)}$    & [fm$^3$]    & $442.461$             & $442.461$             \\ \hline
    $Q_{40}^{(n)}$    & [fm$^4$]    & $6072.06\textbf{9}$   & $6072.06\textbf{8}$   \\ \hline
    $Q_{40}^{(p)}$    & [fm$^4$]    & $5250.20\textbf{1}$   & $5250.20\textbf{2}$   \\ \hline
    $Q_{50}^{(n)}$    & [fm$^5$]    & $-573.0\textbf{47}$   & $-573.0\textbf{37}$   \\ \hline
    $Q_{50}^{(p)}$    & [fm$^5$]    & $573.0\textbf{47}$    & $573.0\textbf{37}$    \\ \hline
    $Q_{60}^{(n)}$    & [fm$^6$]    & $135407.7\textbf{93}$ & $135407.7\textbf{89}$ \\ \hline
    $Q_{60}^{(p)}$    & [fm$^6$]    & $109108.2\textbf{07}$ & $109108.2\textbf{11}$ \\ \hline
    RMS radius        & [fm]        & $5.652$               & $5.652$               \\ \hline
    Charge radius     & [fm]        & $5.686$               & $5.686$               \\
  \end{tabular}
  \caption{Comparison for an HFB calculation of $^{202}$Rn with the D1S Gogny functional. The `1x16' basis uses the deformation and truncation parameters $b_r = b_z = 2.223$fm and $Q = 1.0$. The following constraints are used: $\langle\hat{Q}_{10}\rangle=0\textrm{fm}$, $\langle\hat{Q}_{20}\rangle=3000\textrm{fm}^2$, $\langle\hat{Q}_{30}\rangle=1000\textrm{fm}^3$, $\langle\hat{Q}_{40}\rangle=11322.27\textrm{fm}^4$, $\langle\hat{Q}_{50}\rangle=0\textrm{fm}^5$ and $\langle\hat{Q}_{60}\rangle=244516\textrm{fm}^6$. The convergence target is set to $\epsilon=10^{-8}$.}
  \label{table:202Rn}
\end{table}

Once again, the results are almost identical between both codes. The minor variations between some field energy contributions do not lead to a difference higher than 1keV for the total binding energy. The radii and final multipole moments are quasi-identical. Concerning the performances, HFB3 is more than eleven times faster in total execution time, and needs around 33\% less iterations to converge, compared to HFBTHO. This difference in number of iterations may come from the fact that both codes use different techniques to adjust the Lagrange multipliers associated with the constraints. The HFBTHO code uses the Quadratic Constraint Method~\cite{ring}, while HFB3 uses the Multilinear Constraint Method~\cite{MLCM}. Both codes use a Woods-Saxon pre-conditionner, even if the one used in HFB3 is only a partial implementation (cf. Section \ref{alternatingSolvers}).

The performances shown by HFB3 are quite impressive for these 1-center examples, and calculations for very elongated systems with 2-center HO bases clearly outperform any equivalent 1-center HO basis calculation by any solver on the market.

\section{Contributing}

Unsolicited external contributions to the HFB3 project are more than welcome. Potential contributors should use the official \texttt{GitHub} repository to submit their issues and/or pull requests. The issue tracker is also a good place to discuss the future of the HFB3 project, and its next features.
The HFB3 team is fine with forks, as long as they clearly state that they are derived work from the HFB3 project, and that they respect the licence of HFB3 (\texttt{GPL-v3.0}). If You fork HFB3 and add some nice feature(s), please consider making a pull request to get your feature(s) merged back.

To detect potential regressions, some unit tests are proposed in `\texttt{misc/tests}'. The tests can be built and run with the simple command
\begin{verbatim}
$ bin/run_tests.sh
\end{verbatim}
In a similar way, some unit tests of the Python bindings can be run with the script `\texttt{bin/run\_tests\_python.sh}'.

\section{Conclusion}

This project implements in a modern way a collection of nuclear mean-field techniques ranging from the generation to the post-processing of HFB states. Some of these techniques were only preserved in old Fortran codes, and can now be used again thanks to this project. The full coverage of the Python bindings (100\% of the classes have bindings) allows such a library to be used in modern and complex workflows.

Future versions of HFB3 may include projection techniques (on particle numbers and/or angular momentum) and new nucleon-nucleon effective interactions (DG-type~\cite{Zietek25}). The axial symmetry conservation could also be lifted, enabling the description of triaxial nuclear shapes. The time-reversal symmetry breaking is also a work in progress. The `Link', `Drop' and `Deflation' techniques recently proposed by P.~Carpentier et al~\cite{Carpentier24} may also be included in the HFB3 project, opening the door to exciting new ways to produce continuous ground state and excited PESs, a key ingredient for some future dynamical descriptions of the fission process for example, and also for a new way of practicing spectroscopy studies.

The authors would like to thank the following researchers:
\begin{itemize}
  \item J.-F. Berger for his help with the HFB solver BERGER2CT, which has been the main source of information concerning the 2-center HO basis formalism used in HFB3, and a very accurate source of validation for the implementation of the D1 fields,
  \item M. Girod for his help with the HFB solver AMEDEE, which has been used to validate the implementation of the D2-specific fields in HFB3,
  \item N. Schunck for his help with the HFB solver HFBTHO, which has been used to perform the benchmark presented in this manuscript.
\end{itemize}


\begin{thebibliography}{0}
  \bibitem{ring}P. Ring and P. Schuck, The Nuclear Many-Body Problem (Springer-Verlag, 2000). \url{https://dx.doi.org/10.1063/1.2915762}.
\bibitem{bergerPhd}J.-F. Berger, PhD thesis, Paris XI (1985).
\bibitem{Bender}Bender, M., Rutz, K., Reinhard, P.-G., Maruhn, J.A., 2000. Consequences of the Center-of-Mass correction in nuclear Mean-Field models. EPJ A 7, 467-478. \url{https://doi.org/10.1007/PL00013645}.
\bibitem{Butler}M.N., Sprung, D.W.L., Martorell, J., 1984. An improved approximate treatment of c.m. Motion in DDHF calculations. Nuclear Physics A 422, 157-166. \url{https://doi.org/10.1016/0375-9474(84)90435-4}.
\bibitem{HFBTHOv4}Marevi\'c, P., Schunck, N., Ney, E.M., Navarro P\'erez, R., Verriere, M., O\'Neal, J., 2022. Axially-deformed solution of the Skyrme-Hartree-Fock-Bogoliubov equations using the transformed harmonic oscillator basis (IV) hfbtho (v4.0): A new version of the program. Computer Physics Communications 276, 108367. \url{https://doi.org/10.1016/j.cpc.2022.108367}.
\bibitem{HFBTHOv3}Perez, R.N., Schunck, N., Lasseri, R.-D., Zhang, C., Sarich, J., 2017. Axially deformed solution of the Skyrme-Hartree-Fock-Bogolyubov equations using the transformed harmonic oscillator basis (III) hfbtho (v3.00): A new version of the program. Computer Physics Communications 220, 363-375. \url{https://doi.org/10.1016/j.cpc.2017.06.022}.
\bibitem{HFBTHOv2}Stoitsov, M.V., Schunck, N., Kortelainen, M., Michel, N., Nam, H., Olsen, E., Sarich, J., Wild, S., 2013. Axially deformed solution of the Skyrme-Hartree-Fock-Bogoliubov equations using the transformed harmonic oscillator basis (II) hfbtho v2.00d: A new version of the program. Computer Physics Communications 184, 1592-1604. \url{https://doi.org/10.1016/j.cpc.2013.01.013}.
\bibitem{MLCM}Younes, W., Gogny, D., 2009. Microscopic calculation of 240Pu scission with a finite-range effective force. Phys. Rev. C 80, 054313. \url{https://doi.org/10.1103/PhysRevC.80.054313}.
\bibitem{baranger}Baranger, M. and Veneroni, M., Ann. Phys. 114, 123 (1978). \url{https://doi.org/10.1016/0003-4916(78)90265-8}.
\bibitem{applications} Alamanos, N., Dupuis, M. and Pillet, N., Topical Issue on Finite Range Effective Interactions and Associated Many-Body Methods - A Tribute to Daniel Gogny. Eur. Phys. J. A 53 (2017). \url{https://doi.org/10.1140/epja/i2017-12437-8}.
\bibitem{GognyD1} J. Decharg\'e and D. Gogny, Phys. Rev. C 21, 1568 (1980). \url{https://doi.org/10.1103/PhysRevC.21.1568}.
\bibitem{GognyD1S}J.-F. Berger, M. Girod and D. Gogny, Comp. Phys. Comm. 63, 365 (1991). \url{https://doi.org/10.1016/0010-4655(91)90263-K}.
\bibitem{GognyD1N} F. Chappert, M. Girod and S. Hilaire, Phys. Lett. B, 668 (2008). \url{https://doi.org/10.1016/j.physletb.2008.09.017}.
\bibitem{GognyD1M} S. Goriely, S. Hilaire, M. Girod, and S. P\'eru, Phys. Rev. Lett. 102, 242501 (2009). \url{https://doi.org/10.1103/PhysRevLett.102.242501}.
\bibitem{GognyD2} F. Chappert, N. Pillet, M. Girod, and J.-F. Berger Phys. Rev. C 91, 034312 (2015). \url{https://doi.org/10.1103/PhysRevC.91.034312}.
\bibitem{GognyD3G3} L. Batail, D. Davesne, S. P\'eru, P. Becker, A. Pastore, J. Navarro, Eur. Phys. J. A 59, 219 (2023). \url{https://doi.org/10.1140/epja/s10050-023-01073-w}.
\bibitem{GognyDG} G. Zietek, PhD thesis, Towards a generalized effective nuclear Gogny interaction extended to finite-range spin--orbit and tensor forces, \url{https://theses.hal.science/tel-04394860}, Universit\'e Paris-Saclay (2023).
\bibitem{Ben03}M. Bender, P.-H. Heenen and P.-G. Reinhard, Rev. Mod. Phys. 75, 121 (2003). \url{https://doi.org/10.1103/RevModPhys.75.121}.
\bibitem{Sch16}N. Schunck and L.-M. Robledo, Reports on Progress in Physics 79, 116301 (2016). \url{https://doi.org/10.1088/0034-4885/79/11/116301}.
\bibitem{Schunck19}N. Schunck et al, Energy Density Functional Methods for Atomic Nuclei, IOP Publishing (2019). \url{https://doi.org/10.1088/2053-2563/aae0ed}.
\bibitem{Bro65}C.-G. Broyden, C. G. (1965). `A Class of Methods for Solving Nonlinear Simultaneous Equations'. Mathematics of Computation. 19 (92). American Mathematical Society: 577-593. \url{https://doi.org/10.1090/S0025-5718-1965-0198670-6}.
\bibitem{Rob11}L.-M. Robledo and G.-F. Bertsch, Phys. Rev. C 84, 014312 (2011). DOI: \url{https://doi.org/10.1103/PhysRevC.84.014312}.
\bibitem{WS}R.-D. Woods and D.-S. Saxon, Phys. Rev. 95 (2): 577-578 (1954). \url{https://doi.org/10.1103/PhysRev.95.577}.
\bibitem{OMP}\url{https://en.wikipedia.org/wiki/OpenMP}.
\bibitem{PyPI}\url{https://pypi.org}.
\bibitem{msgp}\url{https://msgpack.org/index.html}.
\bibitem{doxygen}\url{https://www.doxygen.nl}.
\bibitem{Tho22}P. Marevi\'c, N. Schunck, E.M. Ney, R. Navarro P\'erez, M. Verriere and J. O'Neal, Comp. Phys. Comm. 276, 108367 (2022). \url{https://doi.org/10.1016/j.cpc.2022.108367}.
\bibitem{Reg19}D. Regnier, N. Dubray, N. Schunck, Phys. Rev. C 99, 024611 (2019). \url{https://doi.org/10.1103/PhysRevC.99.024611}.
\bibitem{Hilaire07}S. Hilaire and M. Girod, Eur. Phys. J. A 33, 237 (2007). \url{https://doi.org/10.1140/epja/i2007-10450-2}.
\bibitem{Carpentier24}P. Carpentier et al, Phys. Rev. Lett. 133, 152501 (2024). \url{https://doi.org/10.1103/PhysRevLett.133.152501}.
\bibitem{PerezMartin08}S.~Perez-Martin, L.-M.~Robledo, Phys. Rev. C 78, 014304 (2008). \url{https://doi.org/10.1103/PhysRevC.78.014304}.
\bibitem{Gradshteyn65}I.S.~Gradshteyn and I.M.~Ryzhik, Table of Integrals, Series, and Products (Academic Press) (1965). \url{https://doi.org/10.1016/C2010-0-64839-5}.
\bibitem{Delaroche06}J.-P.~Delaroche et al, Structure properties of even-even actinides at normal and super deformed shapes analysed using the Gogny force, Nucl. Phys. A 77, 103, (2006). \url{https://doi.org/10.1016/j.nuclphysa.2006.03.004}.
\bibitem{Zietek25}G.~Zietek et al, Impact of finite-range spin-orbit and tensor terms in Gogny EDF on structure and fission properties, submitted to Phys. Lett. B.
\end{thebibliography}
\end{document}